\documentstyle[sprocl]{article}

\input{psfig}

\def\be{\begin{equation}}
\def\ee{\end{equation}}
\def\bea{\begin{eqnarray}}
\def\eea{\end{eqnarray}}

\begin{document}

\title{COSMOLOGICAL PARAMETERS: \\
do we already know the final answer ?}

\author{MICHAEL ROWAN-ROBINSON}

\address{Astrophysics group, Blackett Laboratory, Imperial College,\\ Prince Consort Rd, London SW7 2BZ\\E-mail: mrr@ic.ac.uk}


\maketitle\abstracts{
Some of the arguments which support the strong concensus for 
an $\Omega_o$ = 0.3, $\lambda_o$ = 0.7 model are reexamined.  Corrections
for Malmquist bias, local flow and metallicity suggest a revised value for
$H_o$ of 63 $\pm$ 6 km$/$s$/$Mpc, improving the age problems for an   
$\Omega_o$ = 1 universe.  The latest CMB results may require a high baryon
density and hence new physics, for example a strong lepton asymmetry. 
Difficulties for the $\Omega_o$ = 1 model
with cluster evolution, the baryon content of clusters, 
and the evidence from Type Ia supernovae favouring low $\Omega_o$, 
$\Lambda > 0$ models, are discussed critically.}

\section{Introduction}

A strong concensus has developed in support of the cosmological parameter set:

$\Omega_o = 0.3, \lambda_o = 0.7, H_o = 75, t_o = 12.6$ Gyr.

Is this justified ?  How certain is the evidence that $\Lambda > 0$ ?  In this
paper I test how strong some of the evidence is, and show that the latest CMB results
may point towards some interesting radical alternatives.

Some of the problems for an $\Omega_0 = 1, \Lambda = 0$ model are:

\begin{itemize}
\item{observed $H_o t_o$ too high}

\item{Type Ia supernovae imply $\lambda_o > 0$ and CMB gives $\Omega_o + \lambda_o = 1$}

\item{cluster abundance evolution favours low $\Omega_o$}

\item{baryon fraction in clusters favours low $\Omega_o$}

\item{P(k) for galaxies requires $\Gamma \simeq \Omega_o h \simeq 0.2$}
\end{itemize}

However these problems may not be as insuperable as they appear.

\section{Hubble constant, $H_{o}$}

New distances from the HST Key program on Cepheids, and their application to 
Type Ia and II supernovae, and the Tully-Fisher, $D_n-\sigma$,
and surface brightness fluctuation methods, have been reviewed by Freedman (2000) and 
Mould et al (2000, and references therein).
In these papers the approach is to determine the Hubble constant via different distance methods
and then combine these estimates statistically.

The philosophy of the present work is that of Rowan-Robinson (1985 (RR85), 1988), 
to form weighted mean distances of groups and clusters 
and then restrict the analysis to high weight distances in order to minimise Malmquist bias

An important ingredient in solving for the Hubble constant is the flow model. 
Estimation of peculiar velocities is one of the largest remaining uncertainties
in $H_o$.  Mould et al (2000) use a simple 3-attractor model.  The PSCz flow model of
Rowan-Robinson et al (2000), based on 88 groups, 854 clusters and 163 voids, 
gives much better resolution of the local flow.

For Type Ia supernovae it is important to correct for internal extinction in the host galaxy.  
Phillips et al (1999) have carried out an important analysis of internal extinction for
Type Ia supernovae and I have used these estimates where available, otherwise the RR85 prescription.
I use the calibration of Gibson et al (2000), based on 8 Type Ia supernovae with HST Cepheid 
distances,  for which the absolute magnitude at maximum light is then $<M_B>$ = -19.47.

An important correction to the HST Cepheid distances is for the differences in metallicity 
between Cepheids in the LMC
and the HST program galaxies.  Mould et al (2000) estimate that this increases the 
HST Cepheid distance scale by 4$\%$.

The HST Key program is based on $\mu_o(LMC) = 18.5$, but Feast (2000) estimates that
this may be too low by 7$\%$, based on Hipparcos parallaxes (after correction for 
metallicity effects in Cepheids).

Figure 1 shows how the mean Hubble constant, calculated for groups and clusters with well-determined 
mean distances, varies with the assumed minimum weight.  Inclusion of lower weight cluster and group
distances leads to an increased mean Hubble constant, almost certainly the effect of Malmquist
bias.  The results are shown for the Mould et al (2000) 3-attractor flow model, and for the
flow model derived from the PSCz analysis (Rowan-Robinson et al (2000).  The difference 
between the flow models is not large, but it is significant.

\begin{figure}
\psfig{figure=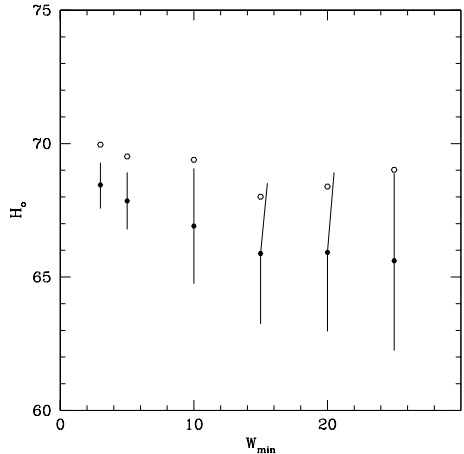,height=2.5in}
\caption{Effect of different minimum weights on mean value of Hubble constant.
Filled circles: PSCz flow model (Rowan-Robinson et al 2000).  Open circles: 3-attractor model
(Mould et al 2000).  No correction for metallicity has been applied.  The error bars
include only statistical uncertainties.}
\end{figure}

Figure 2 shows the Hubble diagram for groups and clusters with distances less than 400 Mpc.
The mean Hubble constant for clusters with weight ( $= \Sigma W/\sigma^2$ ) $>$ 15 is found
to be $H_o = 63 \pm 6$ km$/$s$/$Mpc (Rowan-Robinson, 2000a, in preparation), 
after correction for the effect of metallicity (but with
$\mu_o(LMC) = 18.5$).  I have adopted the uncertainty quoted by Mould et al (2000).

\subsection{Comparison with other estimates}

{\bf Cepheids:}
Freedman (2000) have summarized the results of the HST Key Program using
Cepheids in galaxies out to 20 Mpc to determine the Hubble constant and found
$H_o$ = 74 $\pm$8 km$/$s$/$Mpc.  Mould et al (2000), with their 3-attractor model 
for the local flow, find $H_o$ = 71 $\pm$6 km$/$s$/$Mpc.  The differences between my estimate
and Mould et al's can be assigned as follows:  improved flow model $\Delta H_o$ = -2,
correction for Malmquist bias -2, correction for metallicity -3.

{\bf Supernovae Type Ia:}
Branch (1998) has given an excellent review of the current situation.
Initially widely disagreeing estimates of how the absolute magnitude at maximum light
depends on decay rate have to some extent converged.  Hamuy
et al (1996) analyzed a large sample of nearby supernovae and find
$H_o$ = 63 $\pm$4.5 km$/$s$/$Mpc.  Saha et al (1997) analyzed 7 Type Ia supernovae
with HST Cepheid distances and found $H_o$ = 58 $\pm$8 km$/$s$/$Mpc (61 if they
included a relationship between $M_B$ and $\Delta M_{15}$).  Branch et al (1996)
found $H_o$ = 57 $\pm$4 km$/$s$/$Mpc from a colour matched sample of supernovae and
Tripp (1997) found $H_o$ = 60 $\pm$5 km$/$s$/$Mpc from a sample matched according to
$\Delta m_{15}$.  Riess et al (1996) used a template method in which 
multicolour photometry is used to characterize supernovae in a 1-parameter sequence ('MLCS')
to obtain $H_o$ = 64 $\pm$6 km$/$s$/$Mpc.  

Several groups have brought theoretical models to bear on the determination of
the Hubble constant using Type Ia supernovae.  Hoflich and Khokhlov (1996) compared 
26 supernovae with model light curves and found $H_o$ = 67 $\pm$9 km$/$s$/$Mpc.  Branch (1998)
suggests this should be revised to 56 $\pm$ 5.  The same two authors found $H_o$ = 55
if they included a theoretical version of the  $M_B$ - $\Delta M_{15}$ relation.  
Nugent et al (1995) fitted non-LTE model spectra to observations and found $H_o$ = 60 + 14,-11.
In a recent analysis Tripp and Branch (1999) conclude that the best estimate for the Hubble constant
from Type Ia supernovae was $H_o$ = 62 $\pm$5 km$/$s$/$Mpc. 

\begin{figure}
\psfig{figure=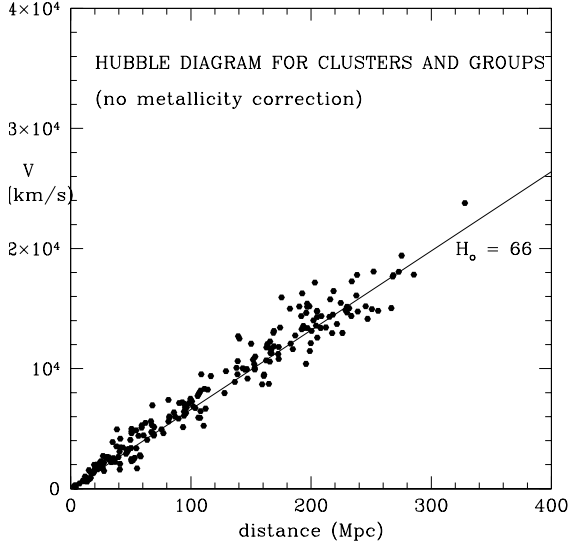,height=3.0in}
\caption{Hubble diagram for groups and clusters with weight $>$ 3.}
\end{figure}

{\bf Gravitational lens time delay:}

An analysis by Falco et al (1997) of the gravitational lens time delay system
0957+561 gave $H_o$ = 62 $\pm$7 km$/$s$/$Mpc.  Koopmans and Fassnacht (1999)
use 5 gravitational lenses to determine $H_o$ = 64 $\pm$11 km$/$s$/$Mpc

{\bf Sunyaev-Zeldovich effect:}
Recent work on SZ clusters includes Myers et al (1997), Birkinshaw (1999) and Reese et al (2000).
Birkinshaw et al (2000) has given $H_o$ = 54 $\pm$ 8 $\pm$ 10 km$/$s$/$Mpc for 9 clusters.

\medskip
To summarize, my estimate for $H_o$, 63 $\pm$ 6 km$/$s$/$Mpc agrees well with independent
estimates.

With $\Omega_o = 1, \lambda_o = 0$, the age of the universe becomes 10.2 $\pm$1.0 Gyr,
while for $\lambda_o = 0.7, \Omega_o = 0.3$, the age becomes 15.0 $\pm$1.4 Gyr.  These
estimates can be compared with the estimate of Chaboyer et al (1998) for the age
of the globular clusters, 11.5 $\pm$ 1.3 Gyr, to which must be added the time
to form the globular clusters, 0.2-2 Gyr, depending on the redshift of formation.
Neither cosmological model would appear to be decisively favoured (or ruled out) by the
age estimate.

\section{CMB and  $ \Omega_{b} h^2$}

The pre-Boomerang and Maxima concensus was that $\Omega_o = 0.3, \lambda_o = 0.7,
\Omega_b h^2 = 0.019 \pm 0.002, h = 0.68, n =1$.  However this provides a
very poor fit to the combined Boomerang and Maxima CMB data.  In particular the second
peak in the angular power spectrum is found to be too weak for this parameter combination.

The concensus fix (Jaffe et al 2000) is to take a higher baryon density, $\Omega_b h^2 = 0.03$.
Unless there is a problem with Big-Bang nucleosynthesis calculations, this would
require the deuterium abundance to be significantly lower than even the low
estimates of Burles and Tytler (1998a,b) (but see Tegmark et al 2000).  It will be interesting
to see whether the weakness of the second peak is confirmed in future observations.

This approach also provides no physical explanation of why the vacuum energy density drops at the
end of inflation by a factor $10^{110}$, only to achieve a significant dynamical role again
at the present epoch.  It is disappointing that quintessence models, which attempt to give some
physical meaning to the evolution of $\Lambda$ do not seem to fit the data as well as a
constant $\Lambda$ does.
  
\section{A radical alternative: lepton asymmetry}

A radical alternative is to consider a strong lepton asymmetry (Harvey and Kolb 1981,
Kang and Steigman 1992, Lesgourgues and Pastor 1999, Kinny and Riotto 1999,
White et al 2000, Lesgourgues and Peloso 2000, Esposito et al 2000).
To maintain charge constancy, this has to be an asymmetry in the neutrino sector.
The nucleosynthesis implications have been explored by Kang and Steigman (1992).
The enhanced relativistic energy-density speeds up the expansion.  Neutrons and
protons decouple earlier, which enhances the abundance of neutrons, but this can
be compensated by increased neutron decay due to an enhanced $\nu_e$ abundance.
To get consistency with the observed primordial light element abundances, we then
need a higher baryonic density.  Lepton asymmetry also has the consequence
(similar to the effects of a positive $\Lambda$ or a hot dark-matter component)
of postponing the epoch of matter-radiation equality, thereby reducing the
parameter $\Gamma$ and improving the fit to the oberved P(k) relative to a
pure CDM scanario.

The lepton asymmetry can be characterized as an effective number of neutrinos, $N_{eff}$.
The ratio of total energy density in extreme relativistic particles to the energy
density in photons, at the epoch of nucleosynthesis is then $\rho_{ER}/\rho_{\gamma}$
= 11$/$4 + 7$/$8 $N_{eff}$.  The expansion speed-up factor after e$\pm$ annihilation 
can be written  $S_o$ = [ 1 + 0.135 ($N_{eff}$ - 3)]$^{1/2}$.
Lesgourgues and Peloso (2000) find a reasonable fit to the Boomerang and Maxima data
with  $N_{eff} = 6, \Omega_o = 1, \lambda_o = 0, \Omega_b h^2 = 0.028, h = 0.70, n = 1$
and $\tau$ (reionization optical depth) = 0.05.  Esposito et al (2000) find that the best fit
to the Boomerang and Maxima data, retaining consistency with observed light element 
abundances, would require  $N_{\nu} = 9 \pm 4 (2-\sigma)$.

However to get a good fit to large-scale structure data, eg P(k), we need

$\Gamma = \Omega_o h/S_o^2 = \Omega_o h/[1+ 0.135(N_{\nu} - 3)] \simeq 0.2$

so for $\Omega_o = 1, h = 0.63$ we need  $N_{\nu} \simeq 19$.

The analysis of Esposito et al (2000) suggest that the best compromise would be
$N_{\nu} = 13$, which would give $\Gamma = 0.26$.  This would give consistency
at the $2-\sigma$ level with the CMB data, light element abundances and large-scale
structure.  

A completely different radical alternative consistent with $\Omega_o = 1, \lambda_o = 0, 
\Omega_b h^2 = 0.019$, and consistent with both the CMB data and large-scale structure data
is the proposal for a phase transition during inflation by Barriga et al (2000).

\begin{figure}
\psfig{figure=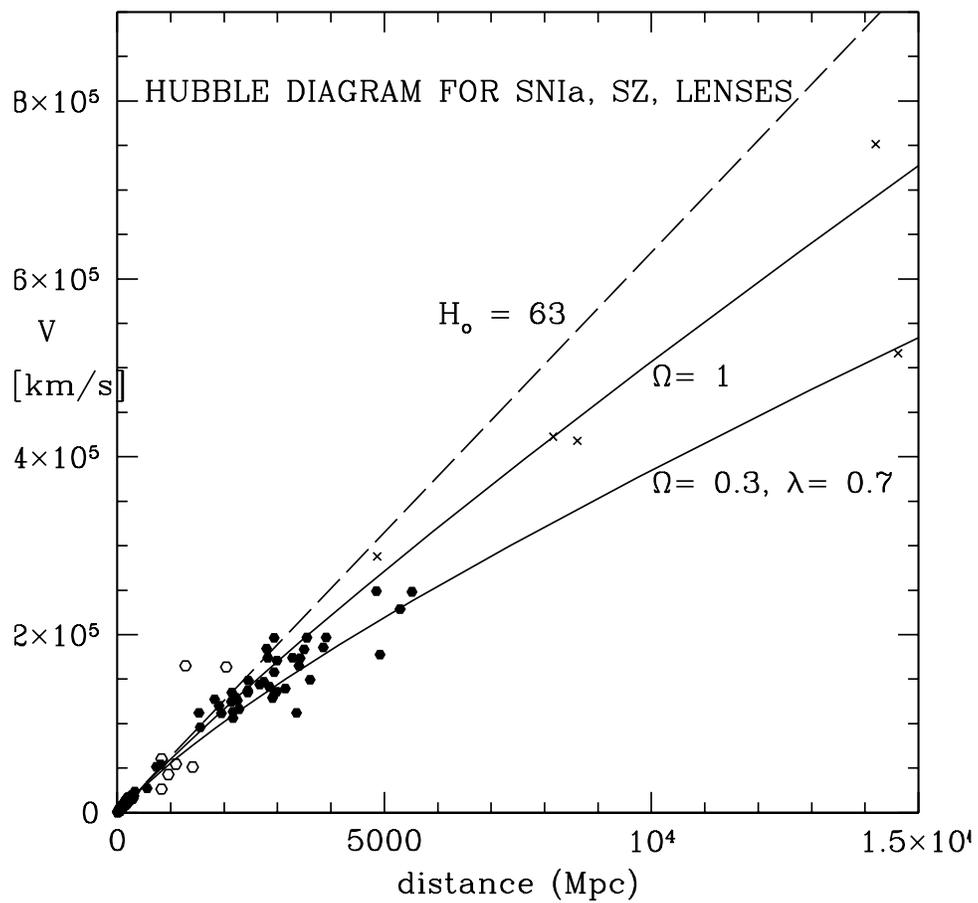,height=5.0in}
\caption{Hubble diagram for SZ clusters (open circles, data from Myers et al 1997, Birkinshaw 1999,
Reese et al 2000), Type Ia supernovae
(filled circles, data from Perlmutter et al 1998) and gravitational lenses 
(crosses, data from Koopmans and Fassnicht 1999).}
\end{figure}

\section{The evidence for positive $\Lambda$ from Type Ia supernovae}

There has been a great deal of excitement about the possibility that Type Ia supernovae
yield positive evidence that $\Lambda >$ 0. 
Recently two groups have published results on the Hubble diagram for Type Ia
supernovae, with over 100 supernovae now discovered at z $>$ 0.3 
(Schmidt et al 1998, Garnavich et al 1998, Riess et al 1998, Perlmutter et al 1999).  
Both groups claim that
models with positive cosmological constant are preferred, and that models with
$\lambda = 0.7, \Omega_{o} = 0.3$ provide the best fit to the data.  
The strength of the signal is that Type Ia supernovae at redshift
0.3-0.9 are about 0.25 magnitudes fainter than local supernoave, if an $\Omega_o$ = 1 
Einstein-de Sitter universe is assumed.  Claims that this is a 7-8 $\sigma$ effect 
therefore depend on a very precise homogeneity of Type Ia supernovae.  The key element
in reducing the scatter in Type Ia supernova absolute magnitudes at maximum light has been the
correlation between absolute magnitude and decline rate ($M_B - \Delta m_{15}$), discussed
above.  If one looks at the paper by Hamuy et al (1996) where this relation is
established for 29 local supernovae, one finds that the situation is not quite
as impressive as has been presented.  It would seem reasonable that to talk about a relationship between
the absolute magnitude at maximum and the decline rate over the next 15 days, it
would be necessary to have detected the calibrating supernovae prior to maximum.
In fact only 10 of the local supernovae were first observed at least one day
before maximum.  For these 10 there is indeed a  $M_B - \Delta m_{15}$ relation, but its
significance is much reduced.  If we derive the calibration from these 10 local
supernovae and apply it to the distant supernovae, the significance of the signal
is reduced from the claimed 7-8 $\sigma$ to only 2-3 $\sigma$,depending on which
calibration is used (Rowan-Robinson 2000b, in preparation).  It appears that the calibrating relation needs to be placed 
on a much stronger basis with nearby supernovae before it can be used to establish
the reality of a cosmological constant.  A good test of homogeneity would be
to find several supernova in a high redshift cluster.

Fig 3 shows the Hubble diagram, velocity versus luminosity distance, 
for snIa, SZ clusters and gravitational lens time delay systems.  The theoretical curves
are shown for an assumed Hubble constant of 63 km$/$s$/$Mpc.  The evidence from snIa supporting
$\Lambda > 0$ comes from supernovae with distances $>$ 3000 Mpc (ie the fainter ones).
The gravitational lenses, reaching to significantly greater redshifts, do not
support $\Lambda > 0$.

There are also some theoretical uncertainties.  Since we do not know for certain whether
nearby supernovae are due to white dwarf deflagration or to white dwarf mergers, there
is the possibility that the proportion of these two types changes with epoch and this could affect
the mean absolute magnitude.
Hoflich et al (1998, 2000) also point out that uncertainties and
evolution of the initial composition in supernovae can have a significant effect on the
determination of cosmological parameters using supernovae.

\section{Other problems: the cluster baryon fraction, cluster abundances}

\begin{itemize}
\item {\bf cluster evolution:} strong negative evolution would be expected in the space density of rich clusters
in an $\Omega_o$ = 1 universe because growth of structure continues to the present day.  This has been claimed
not to have been seen by Bahcall et al (1997), Fan et al (1997), Carlberg et al (1997), Eke et al (1998), 
Bahcall et al (2000).  Bahcall et al (2000) estimate that $\Omega_o = 0.16 \pm 0.05$ and
argue that $\Omega_o$ = 1 is strongly ruled out.  However using an X-ray 
selected sample of clusters, Blanchard et al (2000) estimate $\Omega_o = 0.74 \pm 0.18$, when account is
taken of the cluster temperature distribution function.

\item {\bf the cluster baryon fraction:} Many groups have confirmed the original finding of White et al (1993)
that the baryon content of clusters is too high for an $\Omega_o$ = 1 universe ( White and Fabian (1995), 
Loewenstein and Mushotzky (1996), Mulchaey et al (1996), Evrard (1997)), if the baryon density is taken
to be $\Omega_b h^2 = 0.019 \pm 0.002$.  If however the higher baryonic density implied by the weak second
CMB power spectrum peak seen by Boomerang and Maxima is correct, then the problem is not quite so acute.  For
example Birkinshaw (2000) estimates the $\Omega_b/\Omega_o$ = 0.07, and if this is combined with
$\Omega_b h^2 = 0.028, h = 0.63$, we deduce that $\Omega_o$ = 1.008 !
\end{itemize}

In conclusion there are two viable scenarios consistent with the evidence on cosmological parameters:
an $\Omega_o = 1, \lambda = 0$ universe or a  $\Omega_o = 0.3, \lambda = 0.7$ universe.  Although a number
of lines of evidence, including high redshift Type Ia supernovae, favour the latter, the evidence from
CMB fluctuations may require new physics to resolve.

\end{document}